\documentclass[preprint,12pt]{elsarticle}

\usepackage{graphics}
\usepackage{graphicx}
\usepackage{subcaption}
\usepackage{float}
\usepackage{multirow}
\usepackage{tabulary}
\usepackage[]{algorithm2e}
\usepackage{caption}

\usepackage{amssymb}

\usepackage{lineno}

\journal{Journal Name}

\begin{document}

\begin{frontmatter}



\title{Investigating Causality in Human Behavior from Smartphone Sensor Data: A Quasi-Experimental Approach}


\author{Fani Tsapeli and Mirco Musolesi }

\address{School of Computer Science, University of Birmingham, United Kingdom}

\begin{abstract}
Smartphones have become an indispensable part of our daily life. Their improved sensing and computing capabilities bring new opportunities for human behavior monitoring and analysis. Most work so far has been focused on detecting correlation rather than causation among features extracted from smartphone data. However, pure correlation analysis does not offer sufficient understanding of human behavior. Moreover, causation analysis could allow scientists to identify factors that have a causal effect on health and well-being issues, such as obesity, stress, depression and so on and suggest actions to deal with them. Finally, detecting causal relationships in this kind of observational data is challenging since, in general, subjects cannot be randomly exposed to an event. 

In this article, we discuss the design, implementation and evaluation of a generic quasi-experimental framework for conducting causation studies on human behavior from smartphone data. We demonstrate the effectiveness of our approach by investigating the causal impact of several factors such as exercise, social interactions and work on stress level. Our results indicate that exercising and spending time outside home and working environment have a positive effect on participants stress level while reduced working hours only slightly impact stress. 
\end{abstract}


\end{frontmatter}


\section{Introduction}
Nowadays, people generate vast amounts of data through the devices they interact with during their daily activities, leaving a rich variety of digital traces. Indeed, our mobile phones have been transformed into powerful devices with increased computational and sensing power, capable of capturing any communication activity, including both mediated and face-to-face interactions. User location can be easily monitored and activities (e.g., running, walking, standing, traveling on public transit, etc.) can be inferred from raw accelerometer data captured by our smartphones~\cite{CenceMe,ravi2005activity}. Even more complex information such as our emotional state or our stress level can be inferred either by processing voice signals captured by means of smartphone's microphones~\cite{StressSense,EmotionSense} or by combining information, extracted from several sensors, which correlates with our mood~\cite{Ma2012,Bauer2012,Bogomolov2013,Bogomolov2014}. Moreover, we keep track of our daily schedule by using digital calendars and we use social media to share our experiences, opinions and emotions with our friends.

Leveraging this rich variety of human-generated information could provide new insights on a variety of open research questions and issues in several scientific domains such as sociology, psychology, behavioral finance and medicine. For example, several works have demonstrated that online social media could act as crowd sensing platforms;  the aggregated opinions posted in online social media have been used to predict movies revenues~\cite{Asur2010}, elections results~\cite{Tumasjan2010} or even stock market prices~\cite{Bollen2011}. Social influence effects in social networks have been also investigated in several projects either using observational data \cite{Cha2010,Anger2011} or by conducting randomized trials~\cite{Bond2012,Muchnik2013}. Other works also use mobility traces in order to study social patterns \cite{Onnela2007} or to model the spreading of contagious diseases \cite{Colizza2006}. Moreover, the use of smartphones is increasingly used to monitor and better understand the causes of health problems such as addictions, obesity, stress and depression~\cite{Marsch2012,Pejovic2014}. Smartphones enable continuous and unobtrusive monitoring of human behavior and, therefore,  could allow scientists to conduct large-scale studies using real-life data rather than lab constrained experiments. In this direction, in~\cite{Moturu2011} the authors attempt to explain sleeping disorders reported by individuals, by investigating the correlations between sociability, mood and sleeping quality, based on data captured by mobile phones sensors and surveys. Also, in~\cite{Madan2010} the authors study the links between unhealthy habits, such as poor-quality eating and lack of exercise, and the eating and exercise habits of the user's social network. However, both studies are based on correlation analysis and, consequently, they are not sufficient for deriving valid conclusions about the causal links between the examined variables. For example, an observed correlation between the eating and exercising habits of a social group does not necessarily imply that eating and exercise habits of individuals are influenced by their social group and, therefore, could be modified by changing someone's social group. Instead, the observed correlation could be due to the fact that people tend to have social relationships with people with similar habits.  

The efficient exploitation of human generated data in order to uncover causal links among factors of interest remains an open research issue. Some works have proposed the use of randomized trials \cite{Bond2012,Muchnik2013}. According to this technique, the causal effects of an event or \textit{treatment} are examined by exposing a randomly selected subset of participants (\textit{treatment group}) to this event and comparing the result with the corresponding outcome on a control group (i.e., a subset of participants who have not been exposed to the event). By randomly assigning participants to treatment and control groups it is assured that, on average, there will be no systematic difference on the baseline characteristics of the participants between the two groups. Baseline characteristics are considered to be any characteristics of the subjects that could be related with the study (e.g. in a clinical study the age and the previous health status of the subjects could be considered as baseline characteristics). While randomized trials represent a reliable way to detect causal relationships, they require the direct intervention of scientists in participants' life, which is sometimes unethical or just not feasible. Moreover, such experimental studies cannot exploit the vast amount of observational data that are produced daily. 

Detecting causal relationships in observational data is challenging since subjects cannot be randomly exposed to an event. Thus, subjects that are exposed to a treatment may systematically differ from subjects that are not. In order to eliminate any bias due to differences on the baseline characteristics of exposed and unexposed subjects, scientists need to gather and process information about several factors that could influence the result of the study. There are two main methodologies that can be applied to control such factors: \textit{structural equation modeling}~\cite{mouchart2009structural, spirtes2000causation} and \textit{quasi-experimental designs}~\cite{Campbel1966}. According to the former, the causal effect is estimated using multivariate regression. In detail, the variable representing the causal effect of an event or treatment is regressed using as predictors the variable representing the treatment as well as all the baseline characteristics of the subjects of the study that could influence the result. Structural equation modeling is based on the assumption that the regression model has been correctly specified. False assumptions about the linearity or non-linearity of the model or failure to correctly specify the regression coefficients may result in misleading conclusions. On the other hand, methods based on quasi-experimental designs do not require the specification of a model. Instead, they attempt to emulate randomized trials by exploiting inherit characteristics of the observational data. This can be achieved by comparing groups of \textit{treated} and \textit{control} subjects with similar baseline characteristics (\textit{matching design}).

The purpose of this work is to propose a generic causal inference framework for the analysis of human behavior using digital traces. More specifically, we demonstrate the potential of automatically processing human generated observational digital data in order to conduct causal inference studies based on quasi-experimental techniques. We support our claim by presenting an analysis of the causal effects of daily activities, such as exercising, socializing or working, on stress based on data gathered by smartphones from 48 students that were involved in the StudentsLife project \cite{StudentLife} at Dartmouth College for a period of 10 weeks. The main goal of the StudentsLife project is the study of the mental health, academic performance and behavioral trends of this group of students using mobile phones sensor data. To the best of our knowledge, this is the first work presenting an observational causality study using digital data gathered by smartphones. 

Information about participants' daily social interactions as well as their exercise and work/study schedule is not directly measured; instead we use raw GPS and accelerometer traces in order to infer high-level information which is considered as implicit indicator of the variables of interest. 

No active participation of the users is required, i.e., answering to pop-up questionnaires.
We automatically assign semantics to locations in order to group them in four categories: home, work/university, socialization venues and gym/sports center. By grouping locations into these four categories and continuously monitoring the spatio-temporal traces of users we can derive high-level information as follows:

\begin{itemize}
\item \textbf{Work/University.} By analyzing the daily time that users spend at their workplace we can infer their working schedule. Prolonged sojourn time at work/university could be an indicator of increased workload. 
\item \textbf{Home.} The time that participants spend at home could serve as a rough indicator of their social interactions. Prolonged sojourn time at home could imply limited social interactions or social interactions with a restricted number of people. In general, spending time outside home usually involves some social interaction. An estimation of the total daily time that participants spend at any place apart from their home and working environment could serve as a rough indicator of their non-work-related social interactions. 
\item \textbf{Socialization Venues.} By monitoring users visits at socialization venues such as pubs, bars, restaurants etc, we can infer the time that they spend relaxing and socializing outside home during a day.
\item \textbf{Gym/Sports-center.} Indoor workout can be captured by tracking participants' visits to gyms or sports centers. Outdoor activity can be measured using accelerometer data. 
\end{itemize}

\section{Causal Inference Framework}

Our causality analysis is based on Rubin's counterfactual framework~\cite{Holland86}. According to this framework, a causal problem is formulated as a counterfactual statement which examines what would have been the outcome if an object has been exposed to an event. Since it is impossible to observe for the same object both the result of exposure and non-exposure to an event, causal inference is based on comparing the outcomes on \textit{equivalent} treatment and control groups i.e., treatment and control units with similar baseline characteristics. In this subsection, we discuss a methodology for causal inference in observational data. 

The first step of the analysis is the \textit{description of the variables} of the study. A causality study involves the following variables:
\begin{enumerate}
\item \textit{cause} or \textit{treatment} variable $X$: an independent variable which influences the values of another variable. The treatment variable is usually binary, denoting whether an object of the study has been exposed to a treatment or not. Treatment could be also a discrete variable in case that different levels of treatment are considered. 
\item \textit{effect} or \textit{outcome} variable $Y$: a dependent variable which can be manipulated by changing the variable that represents the cause.
\item a set of $N$ variables $\mathbf{Z} = \{Z^1, Z^2, ..., Z^N\}$, which describes the baseline characteristics of the objects of the study. 
\end{enumerate}

In the second step of the analysis we \textit{define the units} of the study. Each unit corresponds to a set of attributes, derived by the variables of the study, which describe an object (e.g., a person or a thing) on a specific time period. We can use multiple units describing a single object in different time intervals. Thus, a unit $u_{o,t}$ that describes an object $o$ at time $t$ corresponds to a set of values $\{X_{u_{o, t}}, Y_{u_{o, t}}, Z^1_{u_{o, t}}, Z^2_{u_{o, t}}, ..., Z^N_{u_{o, t}}\}$. Given that, in a causation study, the treatment should precede temporally the effect, i.e., the value $X_{u_{o, t}}$ should correspond to the treatment that has been applied to object $o$ before time $t$. In the remainder of the paper, the simplified notation $u$ will be used to describe a unit $u_{o, t}$.

\begin{figure}[!h]
\center
\includegraphics[width=0.35\textwidth]{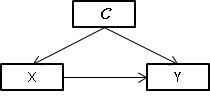}

\caption{Graphical representation of the relationships among the treatment $X$, outcome$Y$ and the set of confounding variables $\mathcal{C}$.}
\label{fig:confounding}
\end{figure}

In order to claim that a value of a variable $Y$ has been caused by a value of a variable $X$ there should be an association between the occurrence of these two values and there should be no other plausible explanation of this association~\cite{Campbel1966}. The first part of this requirement can be examined by performing a simple statistical analysis. However, excluding any other explanation of the observed association is a hard problem since both the treatment and the effect variable may be driven by a third variable. Variables that correlate with both the outcome and the treatment are called \textit{confounding variables} or \textit{confounders}. In Figure~\ref{fig:confounding} we provide a graphical representation of the dependencies between the treatment, outcome and confounding variables. The \textit{identification of the confounders} requires a correlation analysis between each variable $Z^i \in \mathbf{Z}$ and the variables $X$ and $Y$. 

An unbiased causality study requires that the assignment of units to treatments is independent of the outcome conditional to the confounding variables. While in experimental studies this requirement is satisfied by randomly assigning units to treatments, in observational studies we could eliminate confounding bias by comparing units with similar values on their confounding variables but different treatment value (\textit{matching design}). Let us consider a binary treatment $X$, a group of \textit{treated} units $U$ and a group of \textit{control} units $V$ such as $X_u = 1$ $\forall u \in U$ and $X_v = 0$ $\forall v \in V$. Let us also consider a set of confounding variables $\mathbf{\mathcal{C}}$. Ideally, each unit $u \in U$ will be matched with a unit $v \in V$ if $\mathcal{C}^i_u = \mathcal{C}^i_v$,  $\forall \mathcal{C}^i \in \mathbf{\mathcal{C}}$. However, perfect matching is usually not feasible. Thus, treated units need to be matched with the \textit{most similar} control units. Several methods have been proposed to create balanced treated and control pairs~\cite{MatchingSurvey}. After applying a matching method scientists need to check whether the treated and control groups are sufficiently balanced by estimating the standardized mean difference between the groups or by applying graphical methods such as quantile-quantile plots, cumulative distribution functions plots, etc. \cite{Austin2011}. If sufficient balance has not been achieved, the applied matching method needs to be revised.

 Finally, if any confounding bias has been sufficiently eliminated, the treatment effect can be estimated by comparing the effect variable $Y$ of the matched treated and control units. Let us define as $G$ the set of paired treated and control units and $N_G$ the number of pairs. Then, the average treatment effect (ATE) can be estimated as follows:

\begin{equation}
ATE = \frac{\sum_{\forall (u, v) \in G} Y_{u} - Y_{v}}{N_G}
\end{equation}

\begin{figure}[!h]
\includegraphics[width=\textwidth]{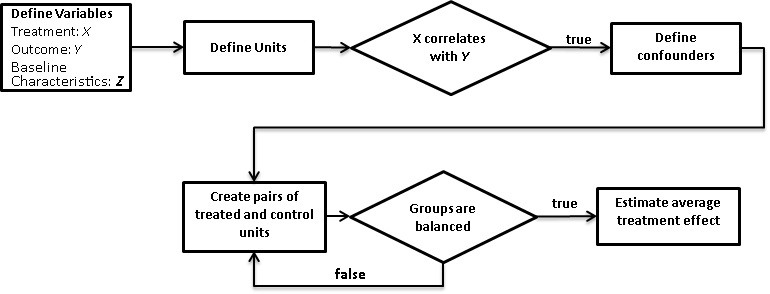}

\caption{Description of the causal inference process in observational data using a quasi-experimental matching design.}
\label{fig:flowChart}
\end{figure}

In Figure~\ref{fig:flowChart}  we provide a graphical representation of the causal inference methodology.

\section{Dataset Description}

The StudentsLife dataset contains a rich variety of information that was captured either through smartphone sensors or through pop-up questionnaires. In this study we use only GPS location traces, accelerometer data, a calendar with the deadlines for the modules that students attend during the term and students responses to questionnaires about their stress level. Students answer to these questionnaires one or more times per day.

We use the location traces of the users to create location clusters. GPS traces are provided either through GPS or through WiFi or cellular networks. For each location cluster, we assign one of the following labels: \textit{home}, \textit{work/university}, \textit{gym/sports-center}, \textit{socialization venue} and \textit{other}. Labels are assigned automatically without the need for user intervention (a detailed description of the clustering and location labeling process is presented at the additional file 1)

We use information extracted from both accelerometer data and location traces to infer whether participants had any exercise (either at the gym or outdoors). The StudentsLife dataset does not contain raw accelerometer data. Instead it provides an activity classification by continuously sampling and processing accelerometer data. The activities are classified to stationary, walking, running and unknown.

We also use the calendar with students' deadlines, which is provided by the StudentsLife dataset, as an additional indicator of students workload. We define as $\mathcal{D}_{deadline}^u$  a set of all days that the student $u$ has a deadline. We define a variable $D^{u, d}$ that represents how many deadlines are close to the day $d$ for a user $u$ as follows:

\begin{equation}
D^{u, d} = 
\left\{
  \begin{array}{rcr}
     \sum_{j} ^{j \in \mathcal{D}_{deadline}^u} \frac{1}{j-d}, & \mbox{ if } j-T_{days}<d<j\\
   
	0, & \mbox{ otherwise}\\
    
\end{array}
\right.
\label{eq:deadlines}
\end{equation}

Thus, $D^{u, d}$ will be equal to zero if there are no deadlines within the next $T_{days}$ days, where $T_{days}$ a constant threshold; otherwise, it will be inversely proportional to the number of days remaining until the deadline. In our experiments we set the $T_{days}$ threshold equal to 3. We found that with this value the correlation between the stress level of the participants and the variable $D^{u, d}$ is maximized. 

Finally, the StudentsLife dataset includes responses of the participants to the Big Five Personality test \cite{Bolger1991}. The Big Five Personality traits describe human personality using five dimensions: \textit{openness},  \textit{conscientiousness}, \textit{extroversion}, \textit{agreeableness}, and \textit{neuroticism}. The personality traits of participants can be used to describe some baseline characteristics of the units and, for this reason, we include them in the study.

\section{Causality Analysis}

We apply the causal inference framework that was previously described in order to assess the causal impact of factors like exercising, socializing, working or spending time at home on stress level. 

\subsection{Variables}

Initially, we define the variables that will be included in the study as follows:

\begin{enumerate}
\item $H^{u, d}_t$: denotes the total time in seconds that the user $u$ spent at home during day $d$ until time $t$;
\item $U^{u, d}_t$: denotes the total time in seconds that the user $u$ spent at university during day $d$ until time $t$;
\item $O^{u, d}_t$: denotes the total time in seconds time that the user $u$ spent in any place apart from his/her home or university during day $d$ until time $t$;
\item  $E^{u, d}_t$: denotes the total time in seconds that the user $u$ spent exercising during day $d$ before time $t$ (it is estimated using both location traces and accelerometer data);
\item $SC^{u, d}_t$: denotes the total time in seconds that the user $u$ spent at any socialization or entertainment venue during day $d$ before time $t$;
\item $S^{u, d}_t$: denotes the stress level of user $u$ that was reported on day $d$ and time $t$. Stress level is reported one or more times per day. Thus, in contrast with the above mentioned variables, $S^{u, d}_t$ is not continuously measured;
\item $PS^{u, d}$: denotes the last stress level that was reported by user $u$ the day $d-1$. This variable remains constant within a day;
\item $D^{u, d}$: represents the upcoming deadlines as described in Equation \ref{eq:deadlines};
\item $E^u$, $N^u$, $A^u$, $C^u$, $O^u$: these five variables denote the extroversion, neuroticism, agreeableness,  conscientiousness and openness of user $u$ based on his Big Five Personality Traits score respectively.
\end{enumerate}

\subsection{Units}

In this study, we examine the effects of five treatments, denoted by the variables $H^{u, d}_{t_i}$, $U^{u, d}_{t_i}$, $O^{u, d}_{t_i}$, $E^{u, d}_{t_i}$ and $SC^{u, d}_{t_i}$ on the stress level of participants, which is described by the variable $S^{u, d}_t$. A unit of the study corresponds to a set of attributes derived by the variables of the experiment. All the variables are sampled every 4 hours, thus there are maximum six samples per day for each participant. Let $T =  \{4 am, 8 am, 12 pm, 16 pm, 20 pm, 24 pm\}$ a set of sampling times and $t_i$ the $i^{th}$ element of $T$. Then a unit corresponds to the set of variables $P^{u, d}_{t_i} =$ ($H^{u, d}_{t_i}$, $U^{u, d}_{t_i}$, $O^{u, d}_{t_i}$, $E^{u, d}_{t_i}$, $SC^{u, d}_{t_i}$, $S^{u, d}_{t_i}$, $PS^{u, d}$, $D^{u, d}$). Since the variable $S^{u, d}_t$ is not continuously measured, it is not feasible to sample it for time $t_i$. Instead, we define $S^{u, d}_{t_i}$ as the average stress level of unit $u$ at day $d$ between time $t_i$ and $t_{i+1}$. Thus,  $S^{u, d}_{t_i}$ is estimated as follows:

\begin{equation}
S^{u, d}_{t_i} = E\{S^{u, d}_t\}, \mbox{ for } t_i \leq t \leq t_{i+1} 
\end{equation}

If there are no stress level reports during this time interval, then the unit that corresponds to the set of variables $P^{u, d}_{t_i}$ will be discarded.

\subsection{Detection of Confounding Variables}

In order to conduct a reliable causation study based on observational data we need to define the confounding variables. While there is a large number of factors that could influence the stress level of participants, the study could be biased only by factors that have a direct influence on both the stress level and the variable that is considered as treatment in the study. Thus, in our case we need to specify factors that could influence both the daily activities of participants and their stress level. For example, the workload of students can influence their activities (e.g., in periods with increased workload some students may choose to change their workout schedule, etc.) and their stress level. Since the workload cannot be directly measured using only sensor data from smartphones, we use as confounding variables other variables that provide implicit indicators of workload such as the time that students spend at home and university and their deadlines. Moreover, participants choice to do an activity may exclude another activity from their schedule and it may also influence their stress level. For example, someone may choose to spend some time in a pub instead of following his/her normal workout schedule. The previous day stress level may also influence both next day's activities and stress level. Finally, several studies have demonstrated that stress level fluctuations are affected by personality traits \cite{Bogomolov2014}. In general, more positive and extrovert people tend to be able to handle stress better than people with high neuroticism score. Moreover, personality characteristics may correlate with the daily schedule that people follow. For example more extrovert people may spend less time at home and more time in social activities. In order to define the covariates of the study we conduct a correlation analysis on the variables of interest. Since the relationship among the variables may not be linear, we apply the Kendall rank correlation. The p-values of the Kendall correlation are presented in Table~\ref{table:kendall_rank_correlation}.

\begin{table}[h!]
  \centering
\begin{tabular}{ |c|c|c|c|c|c|c|}
  \hline 
  {}&{$\mathbf{S^{u, d}_{t_i}}$}&{$\mathbf{H^{u, d}_{t_i}}$}&{$\mathbf{U^{u, d}_{t_i}}$}&{$\mathbf{O^{u, d}_{t_i}}$}&{$\mathbf{E^{u, d}_{t_i}}$}&{$\mathbf{SC^{u, d}_{t_i}}$}\\
  \hline 
  {$\mathbf{H^{u, d}_{t_i}}$}&{0.3557}&{0}&{$6\cdot 10^{-128}$}&{$7\cdot 10^{-182}$}&{0.0161}&{$2.7\cdot 10^{-6}$}\\
  \hline
  {$\mathbf{U^{u, d}_{t_i}}$}&{0.004}&{$6\cdot 10^{-128}$}&{0}&{$2\cdot 10^{-6}$}&{$0.042$}&{0.024}\\
  \hline 
  {$\mathbf{O^{u, d}_{t_i}}$}&{$6\cdot 10^{-5}$}&{$7\cdot 10^{-182}$}&{$2\cdot 10^{-6}$}&{0}&{$10^{-7}$}&{$10^{-13}$}\\
  \hline 
  {$\mathbf{E^{u, d}_{t_i}}$}&{0.0081}&{0.0161}&{0.042}&{$10^{-7}$}&{0}&{0.222}\\
  \hline
  {$\mathbf{SC^{u, d}_{t_i}}$}&{$9\cdot 10^{-5}$}&{$2.7\cdot 10^{-6}$}&{0.024}&{$10^{-13}$}&{0.222}&{0}\\
  \hline   
  {$\mathbf{PS^{u, d}}$}&{$2.7\cdot 10^{-59}$}&{$0.967$}&{$0.0071$}&{0.055}&{0.3897}&{$0.046$}\\
  \hline   
  {$\mathbf{D^{u, d}}$}&{0.024}&{$2.5\cdot 10^{-6}$}&{0.0014}&{0.0018}&{0.002}&{0.0076}\\
  \hline  
  {$\mathbf{E^u}$}&{$1.69 \cdot 10^{-11}$}&{$2.27 \cdot 10^{-5}$}&{$0.059$}&{$4.9\cdot 10^{-4}$}&{$4.1\cdot 10^{-5}$}&{$0.0037$}\\
\hline  
  {$\mathbf{N^u}$}&{$1.81 \cdot 10^{-14}$}&{0.004}&{$1.2 \cdot 10^{-5}$}&{$2.3\cdot 10^{-16}$}&{0.013}&{$6\cdot 10^{-6}$}\\
\hline  
  {$\mathbf{A^u}$}&{0.007}&{0.21}&{$0.15$}&{0.047}&{$0.006$}&{$0.002$}\\
\hline  
  {$\mathbf{C^u}$}&{0.057}&{0.078}&{$0.01$}&{$0.47$}&{0.352}&{0.214}\\
\hline  
  {$\mathbf{O^u}$}&{0.604}&{0.006}&{0.005}&{$2.1 \cdot 10^{-5}$}&{$4.7\cdot 10^{-4}$}&{0.95}\\
\hline  
\end{tabular}
\caption{P-values of Kendall correlation under the null-hypothesis that the examined variables are independent.}

\label{table:kendall_rank_correlation}
\end{table}

Based on these results, the time that students spend at home does not correlate with their stress level. Thus, the variable $H^{u, d}_{t_i}$ will not be included in the causality study. The causal impact of each treatment variable $U^{u, d}_{t_i}, O^{u, d}_{t_i}, E^{u, d}_{t_i}$ and  $SC^{u, d}_{t_i}$ on the effect variable $S^{u, d}_{t_i}$ will be examined using as confounding variables all the variables that correlate with both the treatment and effect based on Table \ref{table:kendall_rank_correlation}. We consider a correlation to be significant enough if the p-value is smaller than 0.1. In Table~\ref{table:confounding}, we present the confounding variables that will be used for each examined treatment. While the variables $O^{u, d}_{t_i}$ and $SC^{u, d}_{t_i}$ are strongly correlated, we do not include $SC^{u, d}_{t_i}$ in the set of confounding variables when the treatment is the variable $O^{u, d}_{t_i}$, since our goal is to study the impact of spending time in any place (including socialization venues) apart from home and working environment.

\begin{table}[h!]
  \centering
\begin{tabular}{ |c|c|c|c|c|c|c|c|c|}
  \hline 
 {\textbf{Treatment}}&\multicolumn{8}{>{\centering}c|}{\textbf{Confounding Variables}}\\
 \hline 
  {$U^{u, d}_{t_i}$ }&{$PS^{u, d}$}&{$D^{u, d}$}&{$O^{u, d}_{t_i}$}&{$E^{u, d}_{t_i}$}&{$SC^{u, d}_{t_i}$}&{$E^u$}&{$N^u$}&{$C^u$}\\
  \hline 
  {$O^{u, d}_{t_i}$ }&{$PS^{u, d}$}&{$D^{u, d}$}&{$U^{u, d}_{t_i}$}&{$E^{u, d}_{t_i}$}&{$E^u$}&{$N^u$}&{$A^u$}&{-}\\
  \hline 
  {$SC^{u, d}_{t_i}$ }&{$PS^{u, d}$}&{$D^{u, d}$}&{$U^{u, d}_{t_i}$}&{$O^{u, d}_{t_i}$}&{$E^u$}&{$N^u$}&{$A^u$}&{-}\\
  \hline
  {$E^{u, d}_{t_i}$}&{$D^{u, d}$}&{$U^{u, d}$}&{$E^u$}&{$N^u$}&{$A^u$}&{-}&{-}&{-}\\
  \hline

\end{tabular}
\caption{Confounding Variables for the different applied treatments.}

  \label{table:confounding}
\end{table}

\subsection{Creation of Treated and Control Groups}

After defining the confounding variables of the study, we need to split the units into control and treatment groups. We consider binary treatments by applying thresholds to the examined treatment variables. Thus, for each of the four examined treatments (i.e., $U^{u, d}_{t_i}$, $O^{u, d}_{t_i}$, $E^{u, d}_{t_i}$, $SC^{u, d}_{t_i}$) the units are split as follows:
\begin{enumerate}
\item $\mathbf{U^{u, d}_{t_i}}$: treatment units are all the units with $U^{u, d}_{t_i}<E\{U^{u, d}_{t_i}\} - \alpha \cdot E\{U^{u, d}_{t_i}\}$ and control all the units with $U^{u, d}_{t_i} \geq E\{U^{u, d}_{t_i}\} + \alpha \cdot E\{U^{u, d}_{t_i}\}$, for a constant $\alpha \in [0, 1)$. Thus, we consider to have a positive treatment value when the university sojourn time is relatively small. 
\item $\mathbf{O^{u, d}_{t_i}}$: treatment units are all the units with $O^{u, d}_{t_i}>E\{U^{u, d}_{t_i}\} + \alpha \cdot E\{U^{u, d}_{t_i}\}$ and control all the units with $O^{u, d}_{t_i} \leq E\{U^{u, d}_{t_i}\} - \alpha \cdot E\{U^{u, d}_{t_i}\}$. Thus, we consider to have a positive treatment value when the time spent in any non-work-related place outside home is relatively large. 
\item  $\mathbf{E^{u, d}_{t_i}}$: treatment units are all the units with $E^{u, d}_{t_i}>0$ i.e. all the units that denote that a user $u$ had some exercise at day $d$ before time $t$. In the control group are units with $E^{u, d}_{t_i}=0$
\item $\mathbf{SC^{u, d}_{t_i}}$: similarly to the treatment variable $E^{u, d}_{t_i}$, treatment units are units with $SC^{u, d}_{t_i} > 0$ and control units with $SC^{u, d}_{t_i} = 0$
\end{enumerate}

Thus, when the treatment variables $U^{u, d}_{t_i}$ and $O^{u, d}_{t_i}$ are considered, units are classified to treated and untreated based on the time they have spent at university or at any place apart from their home and university respectively. However, in order to examine the impact of exercising and visiting socialization venues, the binary treatments are defined by considering only whether there was some exercising activity or a visit to a socialization place or not. We do not study the impact of these factors by considering also the duration of these events since the amount of the data is not sufficiently large. 

Each of the examined treatment variables describes some user behavior or activity from the start of the day to some time $t_i$. Consequently, the comparison of two units with different sampling times $t_i$ is not valid. Thus, we create a group of pairs of treated and control units $G_{t_i}$ for each one of the 6 sampling times $t_i$ such that each treated unit $P^{(u, d)}_{t_i}$ is matched with a control unit $P^{(u, d)'}_{t_i}$ with similar values on its confounding variables. Then, the average treatment effect is estimated as follows:

\begin{equation}
ATE = \frac{\sum_{t_i}\sum_{(P^{(u, d)}_{t_i}, P^{(u, d)'}_{t_i}) \in G_{t_i}} (S_{P^{(u, d)}_{t_i}} - S_{P^{(u, d)'}_{t_i}})}{\sum_{t_i} N_{G_{t_i}}}
\end{equation}

If there is no causal effect of the examined treatment on the stress level then the average treatment effect should be zero. We use a t-test in order to decide whether the observed average treatment effect is statistically significant. 

\subsection{Balance Check}
In order to create balanced treated and control pairs of units we apply the Genetic Matching method~\cite{Diamond2013}. Genetic Matching is a multivariate matching method that applies an evolutionary searching algorithm that estimates weights for each confounding variable in order to achieve an optimal covariates balance. In order to assess if the treated and control pairs are sufficiently balanced, we check the standardized mean difference for each confounding variables of the study. We indicate with $\mathcal{C}$ the set of confounding variables. For each confounding variable $c \in \mathcal{C}$, the standardized mean difference is estimated as follows:

\begin{equation}
SMD_{c} = \frac{\sum_{\forall t_i} \sum_{\forall (P^{(u, d)}_{t_i}, P^{(u, d)'}_{t_i}) \in G_{t_i}}  (c_{P^{(u, d)}_{t_i}} - c_{P^{(u, d)'}_{t_i}})}{\sum_{\forall t_i} N_{G_{t_i}}} / \sqrt{\sigma_c^{T=1}}
\label{eq-standrdized-diff}
\end{equation}

where $\sigma_c^{T=1}$ denotes the variance of the confounding variable $c$ for the treated units. The remaining bias from a confounding variable $c$ is considered to be insignificant if $SMD_{c}$ is smaller than 0.1  \cite{Austin2011}.

\section{Results}

According to Table~\ref{table:kendall_rank_correlation}, extroversion and neuroticism are the two  personality characteristics that strongly correlate with stress level. Thus, we investigate whether some of the examined treatments have a different causal impact on people with high extroversion or neuroticism scores. We conduct our study \textit{collectively} for the whole population and \textit{selectively} for people with high extroversion score and people with high neuroticism score. We define as \textit{Extroverts} all the participants with extroversion score larger than the average extroversion score of all the participants. Correspondingly, we define a subpopulation of \textit{Neurotics} composed of participants with neuroticism score higher than the average score of the population.

\begin{figure}[t!]
\centering
        \begin{subfigure}[b]{0.5\textwidth}
                \includegraphics[width=\textwidth]{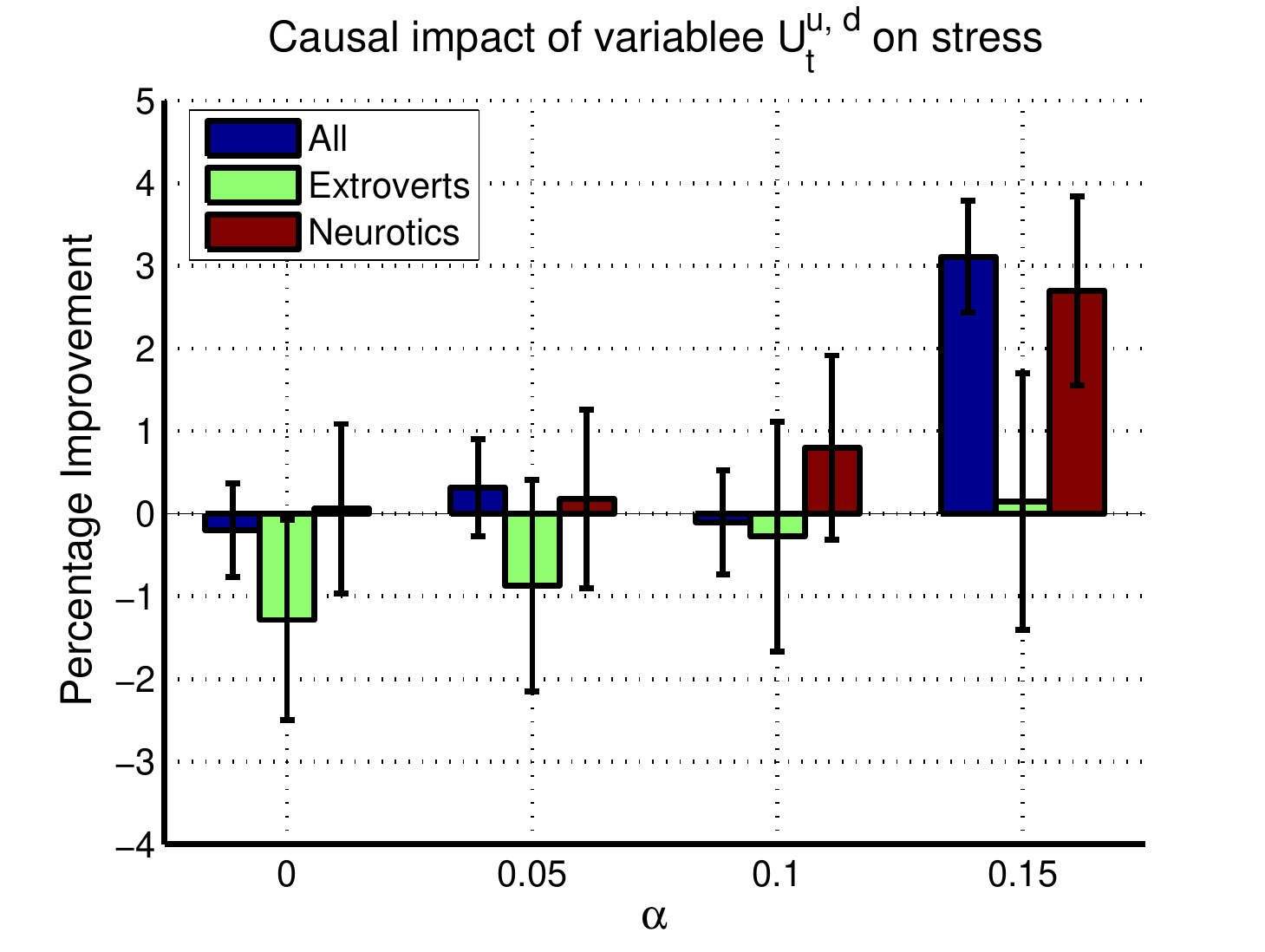}
                \caption{Causal effect of variable $U^{u, d}_{t_i}$ on the stress level for the four examined thresholds}
                \label{fig:uni}
        \end{subfigure}%
~
        \begin{subfigure}[b]{0.5\textwidth}
                \includegraphics[width=\textwidth]{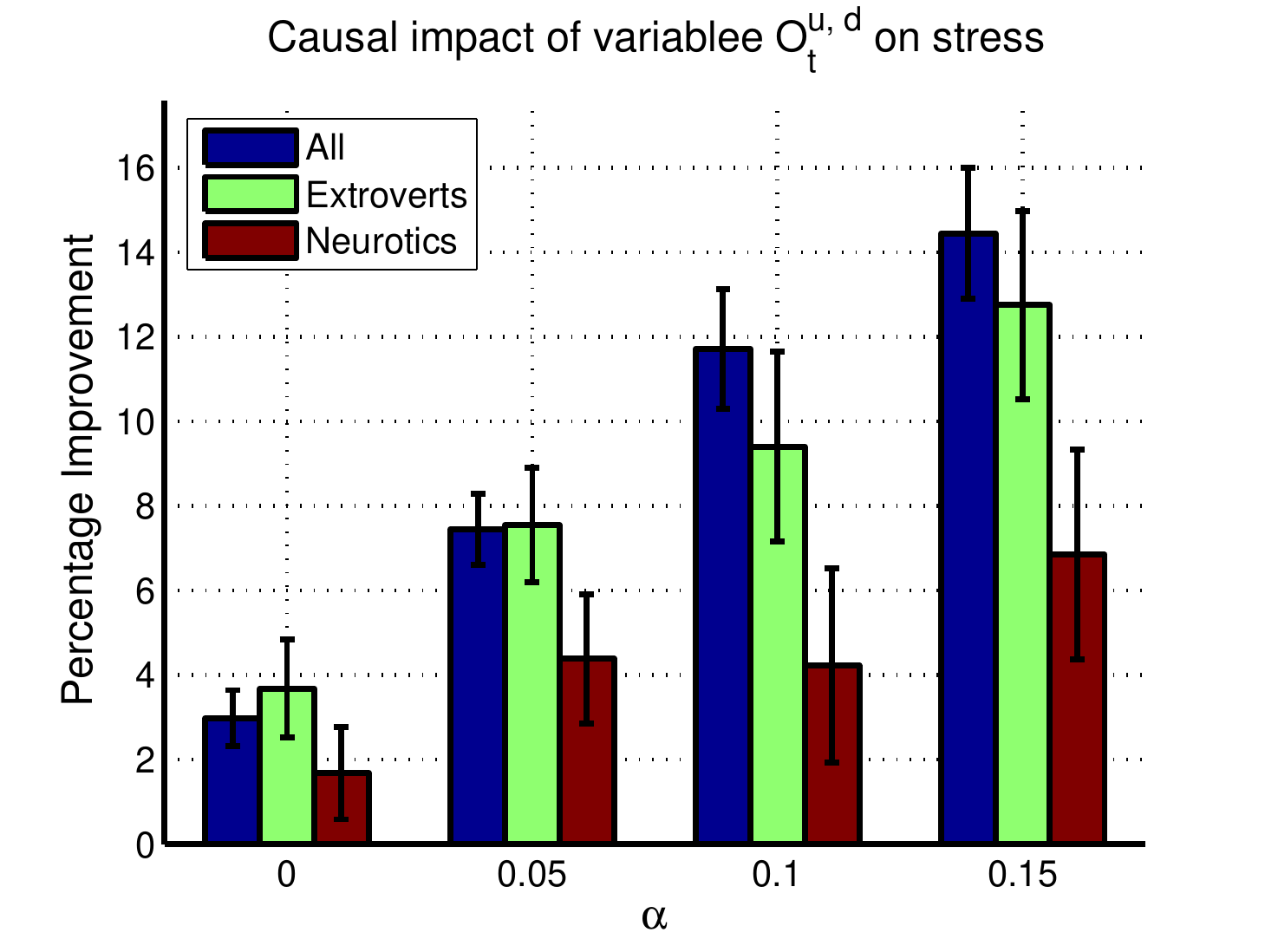}
                \caption{Causal effect of variable $O^{u, d}_{t_i}$ on the stress level for the four examined thresholds}
                \label{fig:out}
        \end{subfigure}
~
 \begin{subfigure}[b]{0.5\textwidth}
                \includegraphics[width=\textwidth]{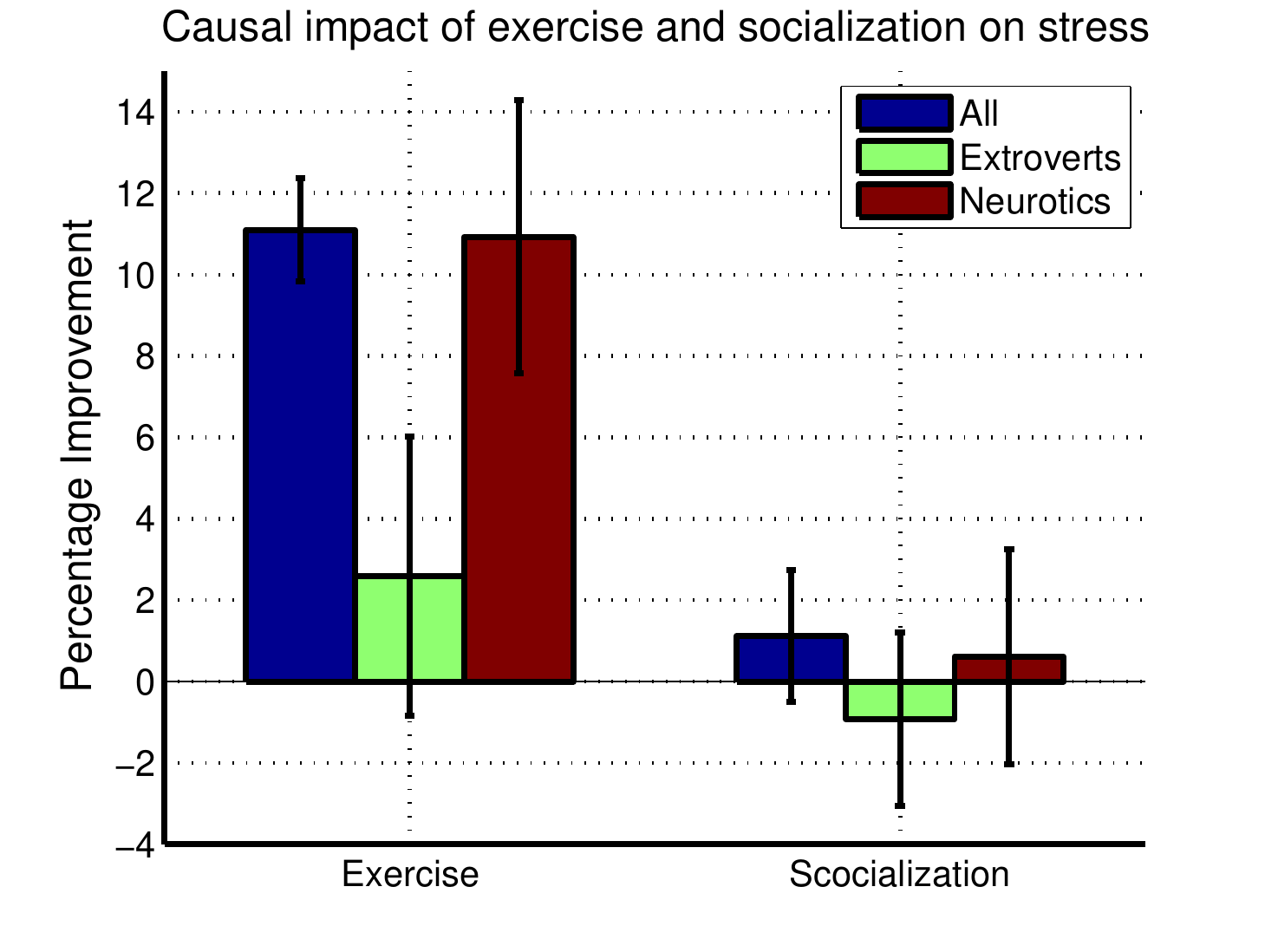}
                \caption{Causal effect of variables $E^{u, d}_{t_i}$ and $Sc^{u, d}_{t_i}$ on the stress level }
                \label{fig:ex-soc}
        \end{subfigure}
\caption{Percentage improvement on the stress level of treated units compared to control units when each one of the examined treatments is applied. Percentage improvement is estimated as $\frac{ATE}{E\{S^{(u, d)'}_{t_i}\}}\times 100$. }
\label{fig:causality}
\end{figure}

\begin{figure}[h]
\centering
        \begin{subfigure}[b]{0.5\textwidth}
                \includegraphics[width=\textwidth]{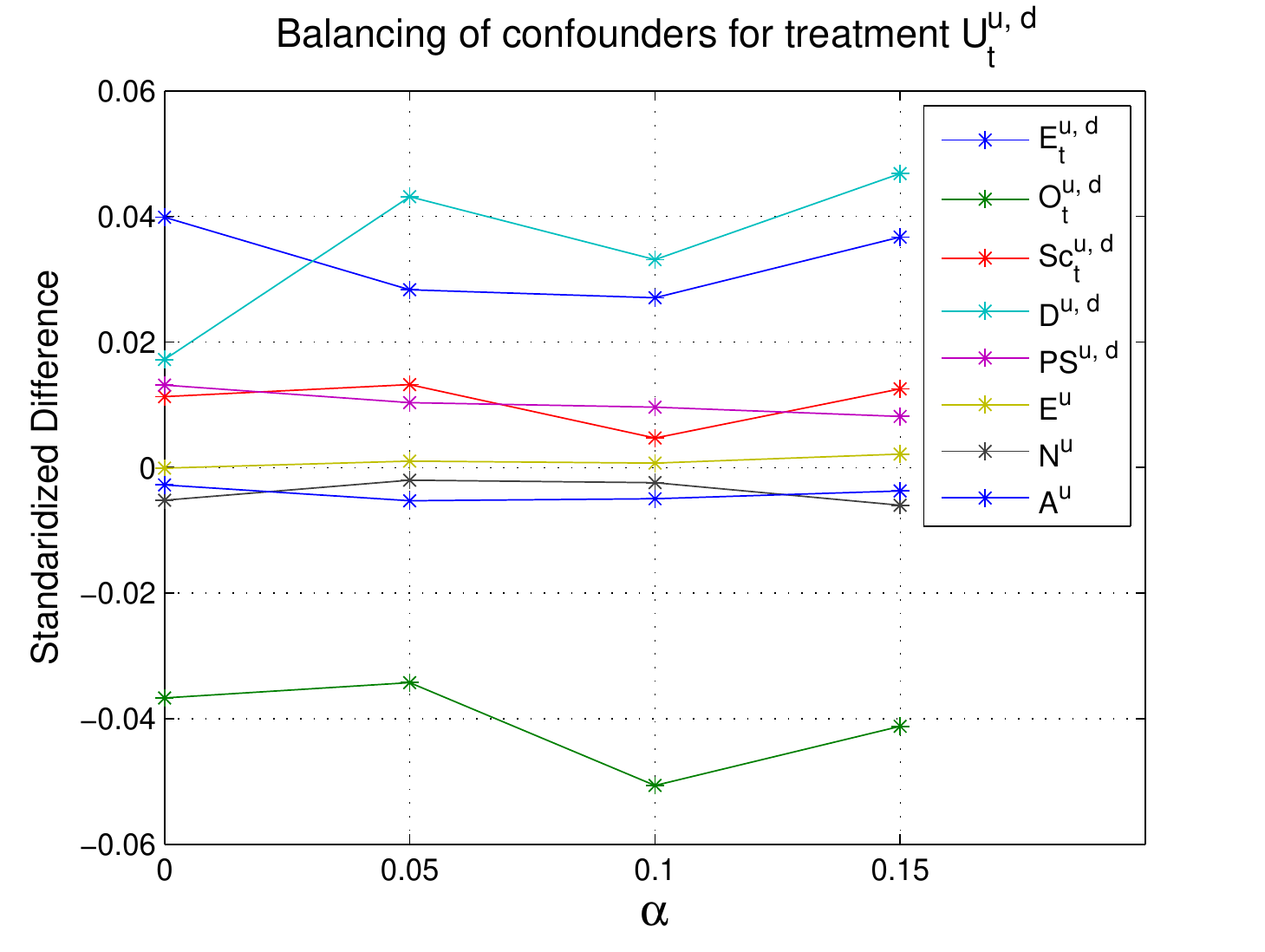}
                \caption{}
                \label{fig:uni-balance}
        \end{subfigure}%
~
        \begin{subfigure}[b]{0.5\textwidth}
                \includegraphics[width=\textwidth]{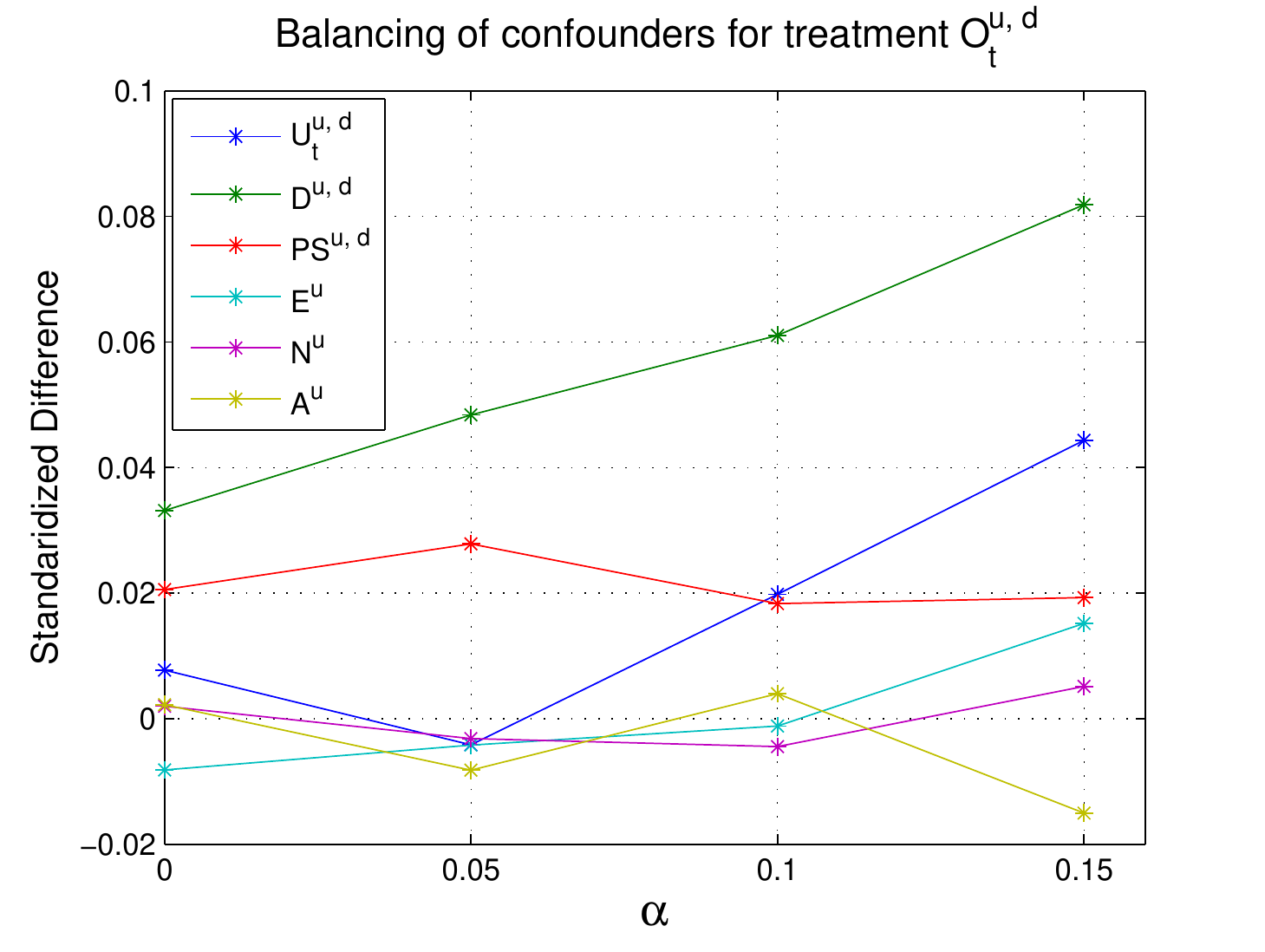}
                \caption{}
                \label{fig:out-balance}
        \end{subfigure}
\caption{Standardized difference between treated and control samples for each confounding variable when the applied treatment is (a) the variable $U^{u, d}_{t_i}$ and (b) the variable $O^{u, d}_{t_i}$. The standardized difference for all the confounding variables is less than 0.1, thus the groups are balanced.}
\label{fig:balancing}
\end{figure}

In Figure \ref{fig:causality}, we present the average treatment effect (ATE) normalized by the average stress level of the control units along with the 95\% confidence intervals for each one of the four examined treatment variables. For the treatment variables $U^{u, d}_{t_i}$ and $O^{u, d}_{t_i}$ we present results for $\alpha$ equal to $0$, $0.05$, $0.1$ and $0.15$. We do not present results for larger $\alpha$ values since the number of samples that are discarded is large and the remaining data are not sufficient for statistically significant conclusions. In Figure~\ref{fig:balancing} and Table \ref{tab:balancing}, we present the standardized difference, as described in Equation \ref{eq-standrdized-diff}, for all the confounding variables that were used in each one of the causation studies. According to our results, the standardized difference between treated and control samples is smaller than 0.1 for all the confounding variables thus any confounding bias has been sufficiently minimized. 

Our results indicate that the time that students spend at university has only a weak causal impact on the stress level when participants' samples are split into treatment and control groups using an $\alpha$ value equal to $0.15$. In detail, participants report 3.1\% (with confidence interval $\pm 0.7$\%)  lower stress level the days that their sojourn time at university is 15\% lower than the average university sojourn time of the whole population compared to days that the university sojourn time is 15\% larger than usual. However, when the analysis is limited to people with high extroversion score, there is no statistically significant evidence that the time that students spend at university has any causal effect on stress. When smaller $\alpha$ values are considered, the causality score is close to zero for the examined set of students as well as for the \textit{Extroverts} and \textit{Neurotics} sub-populations.

Based on our results, the time that students spend in any place apart from their home and university has a significantly strong causal impact on their stress level. As depicted in Fig. \ref{fig:causality}.b, students have reported around 3\% (with confidence interval $\pm 0.65$\%) lower stress level the days that they spend more time outside than the average time compared to days that they spend less time outside (i.e., $\alpha = 0$), when the whole set of participants is considered. Similar results are observed when the \textit{Extroverts} and \textit{Neurotics} sub-populations are examined (the observed difference is not statistically significant given the 95\% confidence intervals of the study). When the value of $\alpha$ is increased, the causal impact of the examined variable is stronger. For $\alpha = 0.15$, the improvement on the stress level for students who spend more time outside is 14.45\% (with confidence interval $\pm 1.5$\%) when the total population is considered. The results are similar when the study is limited to students with high extroversion score. However, the examined variable has a significantly lower impact on stress level when the sub-population of \textit{Neurotics} is considered. 

In Fig. \ref{fig:causality}.c, we examine the impact of exercising or visiting socialization venues on stress level. While the variable $SC^{u, d}_{t_i}$  is strongly correlated with the stress level, according to our results, there is no causal link between them. This indicates that, while people benefit from spending time outside home or working environment in general, there is no statistically significant benefit from visiting specific venues. Finally, exercising has positive effect on the stress level of the examined population. However, results are different when the  \textit{Extroverts}  and \textit{Neurotics} sub-populations are examined separately. Exercising has a stronger positive effect on the stress level of participants with high neuroticism score while there is no statistically significant benefit for people with high extroversion score.  

\begin{table}[h!]
  \centering
\begin{tabular}{ |c|c|c|c|c|c|c|c|}
  \hline 
  {}&{$PS^{u, d}$}&{$D^{u, d}$}&{$U^{u, d}_{t_i}$}&{$O^{u, d}_{t_i}$}&{$E^u$}&{$N^u$}&{$A^u$}\\
  \hline 
  {$SC^{u, d}_{t_i}$ }&{$-0.0035$}&{$0.0442$}&{$0.0046$}&{$-0.0148$}&{$-0.0069$}&{$-0.0065$}&{$0.0001$}\\
  \hline
  {$E^{u, d}_{t_i}$}&{-}&{$0087$}&{$-0.0011$}&{-}&{$0.0047$}&{$0$}&{$0.0043$}\\
  \hline
\end{tabular}
\caption{Standardized difference between treated and control samples for each one of the confounding variables when the applied treatments correspond to the variables $SC^{u, d}_{t_i}$  and $E^{u, d}_{t_i}$.}

  \label{tab:balancing}
\end{table}

\section{Discussion}

In this work, we presented a framework for detecting causal links on human behavior using mobile phones sensor data. We have studied the causal effects of several factors, such as working, exercising and socializing, on stress level of 48 students using data captured by smartphones sensors. Our results suggest that exercising and spending time outside home or university have a strongly positive causal effect on participants stress level. We have also demonstrated that the time participants stay at university has a positive causal impact on their stress level only when it is considerably lower than the average daily university sojourn time. However, this impact is not remarkable. 

Moreover, we have observed that some of the examined factors have different impact on the stress level of students with high extroversion score and on students with high neuroticism score. More specifically, more extrovert students benefit more from spending time outside home or university while more neurotic students benefit more from exercising. 

Our study mainly relies on raw sensor data that can be easily captured with smartphones. We have demonstrated that information extracted by simply monitoring users' location and activity (through accelerometer) can serve as an implicit indicator of several factors of interest such as their working and exercising schedule as well as their daily social interactions. Inferring this high-level information using raw sensor data instead of pop-up questionnaires has three main advantages: 1) it offers a more accurate representation of participants activities over time since data are collected continuously, 2) data are collected in an obtrusive way without requiring participants to provide any feedback; this minimizes the risk that some users will quit the study because they are dissatisfied by the amount of feedback that they need to provide, 3) data gathered through pop-up questionnaires may not be objective since participants may provide either intentionally or unintentionally false responses. On the other hand, inferences based on sensor data could also be inaccurate either due to noisy sensor measurements or due to the fact that the variable of interest is inferred by the sensed data rather than directly measured. For example, in our case we assume that a visit to a sports center implies that the user had some exercise. However, the user may have visited this place to attend a sports event or just to meet friends. Assessing the degree of uncertainty that information inference from sensor measurements involves and incorporating this uncertainty into the causation study represents an interesting research area for further investigation.

Finally, this study involves a limited number of participants who do not constitute a representative sample of the population; therefore extrapolating general conclusions about the causal impact of the examined factors on stress level is not feasible. However, the purpose of this article is to demonstrate the potential of utilizing smartphones in order to conduct large-scale studies related to human behavior, rather than present a thorough investigation on factors that influence stress.

\appendix

\section{Location Clustering and Labeling}
We create location clusters using raw GPS traces. In order to increase the accuracy on location estimation we consider only GPS samples with accuracy less than 50 meters. Moreover, we ignore any samples that were collected while the user was moving. For each new GPS point, we create a cluster only if the distance of this point with the centroid of any of the existing clusters is more than 50 meters. Otherwise, we update the corresponding cluster with the new GPS sample. Every time a new GPS sample is added to a cluster, the centroid of the cluster is also updated. The pseudo code of the location clustering algorithm is presented at Algorithm 1. 

\begin{algorithm}[!t]
 \KwData{Set of location points $L = \{l_1, l_2, ..., l_n\}$}
 \KwResult{ Set of Clusters $C = \{c_1, c_2, ..., c_m\}$}
 $C:=\{\}$\; 
\For{each l $\in$ L}{
  \If {accuracy(l)$>$50}{
     continue\;
   }
   $locationClusteredFlag := 0$\;
   \For{each c $\in$ C}{
	 $H:=\{Z^{j,k}: Z^{j,k} \in P\}$\;
	 \If{distance(l, centroid(c))$<$50}{
        $c:=c \cup \{l \}$\;
	  $locationClusteredFlag := 1$\;
	  break\;
   }	
  }
\If{locationClusteredFlag = 0}{
	$newCluster:=\{l\}$\;
      $C:=C \cup \{newCluster\}$\;
}
 }
 \caption{Location clustering }
 \label{algocluster}
\end{algorithm}

Each location cluster is labeled as \textit{home}, \textit{work/university}, \textit{gym/sports-center}, \textit{socialization venue} or \textit{other}. The label \textit{socialization venue} is used to describe places like pubs, bars, restaurants and cafeterias. The label \textit{other} is used to describe any place that does not belong to the above mentioned categories. We label as \textit{home} the place that people spend most of the night and early morning hours. In order to find clusters that correspond to gyms/sports-centers or socialization venues we use the Google Maps JavaScript API \cite{GoogleAPI}. Google Maps JavaScript API enable developers to search for specific type of places that are close to a GPS point. The type of place is specified using specific keywords from a list of keywords provided by this API. We use the centroid of each unlabeled cluster to search for nearby places of interest. Places that correspond to \textit{gym/sports centers} are specified by the keyword \textit{gym} and places that correspond to socialization venues are specified by the keywords \textit{bar}, \textit{cafe}, \textit{movie\_theater}, \textit{night\_club} and \textit{restaurant}. For each unlabeled cluster we conduct a search for nearby points of interests. If a point of interest with distance less than 50 meters from the cluster centroid is found, we label the cluster as \textit{gym/sport-center} or \textit{socialization venue} depending on the point of interest type. Otherwise the cluster is labeled as \textit{other}. Any place within the university campus that is not labeled as \textit{gym/sport-center} or \textit{socialization venue} is labeled as \textit{work/university}.

\section{Matching Method}

For matching the treatment and control units we use the MatchIt R package~\cite{ho1737matchit} which includes an implementation of the genetic matching algorithm described above. Several optimization criteria can be used with Genetic Matching~\cite{sekhon2008multivariate}. Here, the balance metric that the Genetic Matching algorithm optimizes is the mean standardized difference of all the confounding variables. We use matching with replacement, i.e., each control unit can be matched to more than one treatment units. Matching with replacement can reduce the bias since control units which are very similar to treatment units can be exploited more. We use a matching ratio equal to 2. This means that each treatment unit will be matched with up to 2 control units.






\end{document}